\def\A/{${\mathcal A}$}
\def\U/{${\mathcal U}$}

\newtheorem{thm}{Theorem}
\newtheorem{defn}[thm]{Definition}
\newtheorem{lem}[thm]{Lemma}

\documentclass[10pt]{IEEEtran}

\usepackage{amssymb}
\usepackage{amsmath}
\usepackage{algorithmic}

\begin{document}

\title{Multilevel Threshold Secret and Function Sharing based on the Chinese Remainder Theorem}

\author{O\u{g}uzhan~Ersoy, Kamer~Kaya and~Kerem~Ka{\c{s}}kalo{\u{g}}lu}

\maketitle

\begin{abstract}
A recent work of Harn and Fuyou presents the first
multilevel (disjunctive) threshold secret sharing scheme based on the
Chinese Remainder Theorem. In this work, we first show that the proposed
method is not secure and also fails to work with a certain natural
setting of the threshold values on compartments. We then propose a
secure scheme that works for all threshold settings. In this scheme, we
employ a refined version of Asmuth-Bloom secret sharing with a special
and generic Asmuth-Bloom sequence called the {\it anchor sequence}.
Based on this idea, we also propose the first multilevel conjunctive
threshold secret sharing scheme based on the Chinese Remainder Theorem.
Lastly, we discuss how the proposed schemes can be used for multilevel
threshold function sharing by employing it in a threshold RSA
cryptosystem as an example.
\end{abstract}

\begin{IEEEkeywords}
Secret sharing, multilevel function sharing, multilevel threshold cryptography,   Chinese Remainder Theorem.
\end{IEEEkeywords}

\section{Introduction}

The concept of secret sharing is being used in many cryptographic
protocols. As independently proposed by Shamir~\cite{Shamir79} and
Blakley~\cite{Blakley79}, a {\em secret-sharing scheme}~(SSS) involves a
\textit{dealer} who has a \textit{secret} $s$, a set of
\textit{participants} \U/ that the secret is shared amongst, and a
collection \A/ of the authorized subsets of the \U/ which is
called the \textit{access structure}. In a SSS, the dealer distributes the
shares to the participants such that only the subsets in \A/ can
reconstruct the secret from the corresponding shares. Furthermore, a SSS
is called \textit{perfect} if all the subsets not in \A/ will have the
same probability of guessing the secret as if they had no shares. We
refer the reader to a comprehensive survey~\cite{Beimel11} for practical
applications of secret sharing such as building authentication protocols which stay secure
even under the leakage of a number of servers' data. 

In {\em threshold secret sharing}, the access structure is defined by a
threshold on the cardinality of authorized subsets: a $(t,n)$-SSS refers
to a scheme in which any $t$ out of $n$ participants can recover the
secret. Apart from Shamir's Lagrange interpolation-based
scheme~\cite{Shamir79} and Blakley's scheme utilizing the idea that any
$n$ nonparallel ($n-1$)-dimensional hyperplanes intersect at a specific
point~\cite{Blakley79}, Chinese Remainder Theorem~(CRT)-based threshold schemes by
Mignotte~\cite{Mignotte83} and Asmuth and Bloom~\cite{Asmuth83} also
exist. While Mignotte's $(t,n)$ scheme is not perfect in the sense that
less than $t$ shares reveal information about the secret, Asmuth-Bloom's
scheme attains a better security level with a careful choice of
parameters. We refer the reader to~\cite{Quisquater02} for an extensive
study on the security of CRT based SSSs.

Given the universal participant set \U/, a partition of \U/ into
disjoint subsets, i.e., {\em compartments}, is used to define a {\em
multipartite access structure} on \U/. Unlike traditional threshold
secret sharing that has only one threshold, different thresholds and conditions may be imposed for
different compartments. On the other hand, multipartite schemes do not
distinguish the members of the same compartment. 

Although there exist methods for general access structures,
e.g.,~\cite{Ito87,Iftene07,Bozkurt10}, the schemes designed for specific
access structures are almost always more efficient and hence, more
practical. Such an access structure which has applications in practice
is the {\em multilevel/hierarchical access structure}, a special form of
the multipartite case, that employs a hierarchy between the compartments
where the members of a superior compartment are more powerful and can
replace the participants of an inferior one following the hierarchy
definition of Simmons~\cite{Simmons90} that is further studied
in~\cite{Ghodosi98}. Simmons gave the following example: assume 
that a bank transfer requires authorization and any two vice
presidents or any three senior tellers are authorized to approve. In
this example, there are two compartments~(vice presidents and senior
tellers) where the members of the former can also replace members of the
latter. That is, a vice president together with two senior tellers are
able to approve the transfer as well. In a recent work of Harn and
Fuyou~\cite{Harn14}, a multilevel CRT-based SSS is proposed for an
access structure involving a hierarchy of compartments as in the definition
of Simmons.

The above mentioned access structure is {\it disjunctive}; if a
coalition satisfies any of the threshold conditions of the compartments
(that is, the presence of either two vice presidents {\bf or} three senior
tellers), then it is in the access structure. A more restricted {\it conjunctive} form
where all the compartments' thresholds need to be satisfied by a
valid coalition can also be employed in practice. For example, suppose a
bank transfer now requires the authorization of any two vice presidents
{\bf and} any three senior tellers for the example above. Note that different
from the disjunctive scheme, a coalition needs to
satisfy all the thresholds in the conjunctive form. Hence, with this
requirement, a vice president and two senior tellers~(or only three
senior tellers) cannot authorize a transfer as they could in the
disjunctive case. Although a CRT-based conjunctive threshold SSS has
been proposed by Iftene~et~al.~\cite{Iftene05}, to the best of our
knowledge, there is no hierarchical/multilevel conjunctive secret
sharing scheme based on CRT in the literature.

There are multiple contributions of this paper: First, we show that the
Harn-Fuyou scheme cannot be applied~(i.e., is not well-defined) for all
the access structures \A/ in the multilevel setting, and furthermore it
is not secure, i.e., the secret can be reconstructed by an unauthorized
coalition that is not in \A/. Second, by using an {\it anchor
Asmuth-Bloom sequence}, we propose a more naive and novel CRT-based SSS for the multilevel
access structures which does not suffer from these drawbacks. And third,
based on similar techniques, we propose the first multilevel conjunctive
threshold SSS based on CRT.  

In addition to the main contributions, we also discuss  how the proposed schemes can
be employed for \textit{multilevel function sharing}; a natural
extension of secret sharing. A plain SSS is inappropriate for public key
cryptography, since when a (shared) secret is reconstructed, it is known
by all the participants and cannot be used again. A {\em function sharing scheme}~(FSS) employs
a SSS to share the keys/secrets so that operations such as decryption
or signing can only be performed by a valid coalition (in \A/) and without
revealing the secret. As usual, a coalition that is not in \A/ cannot perform
such operations and cannot obtain any information on the secret.
Function sharing schemes enable many applications in practice such as
the fair sale of digital content in exchange for digital receipts, secure
bidding, and secret election protocols. There are numerous studies on
function sharing; the work of Shoup~\cite{Shoup00} can be considered as
one of the milestones of the field, proposing a provably secure,
non-interactive FSS joining RSA and Shamir's SSS. The first CRT based
function sharing schemes for RSA, ElGamal and Paillier cryptosystems are given
in~\cite{Kaya07}.

After covering some preliminary definitions and schemes in
Section~\ref{sec:prel}, we point out some shortcomings and the insecurity of
the Harn-Fuyou scheme in Section~\ref{sec:HY}. We present our
conjunctive and disjunctive multilevel secret sharing schemes in
Section~\ref{sec:MSSS}, and in Section~\ref{sec:MFSS}, we
discuss and show how the proposed schemes can be adopted for function
sharing by using RSA signature/decryption~\cite{Rivest78} as an example.
Section~\ref{sec:conc} concludes the paper. 

\section{Background and Preliminaries}\label{sec:prel}

\noindent Given the following system of congruences
\begin{align*}
x=&s_1 \, mod\, p_1\\
x=&s_2 \, mod\, p_2\\
&\vdots\\
x=&s_n \, mod\, p_n,
\end{align*}
\noindent the Chinese Remainder Theorem states that there is a unique
solution $x\in \mathbb{Z}_P$ such that $$x
= \sum_{i=1}^n \frac{P}{p_i} I_i s_i \bmod P,$$ where
$P = lcm(p_1, p_2, \ldots,
p_n)$ and $I_i$ is the inverse of $P/p_i$ in modulo $p_i$, i.e., $\frac{P}{p_i}I_i \, \bmod \, p_i = 1$. 
Thus when the $p_i$ values are chosen pairwise coprime~(or all prime)
$P$ becomes
$p_1 p_2 \ldots p_n$. 

\subsection{Mignotte's secret sharing}
Mignotte's SSS is a direct application of CRT with one specification:
With $n$ participants and a threshold $t \leq n$, given the sequence of
pairwise coprime positive integers~(or primes) $p_1<p_2<\ldots<p_n$, the secret
$s$ is chosen from the interval $(p_{n-t+2}p_{n-t+3}\ldots p_n \,,\, p_1p_2 \ldots p_t)$. 
The share of each participant $u_i$ is $s_i=s \mod 
p_i$. Since $s$ is greater than the product of the greatest $t-1$ primes,
a set of $t-1$ participants cannot (uniquely) reconstruct the
secret. On the other hand, $t$ or more participants can reconstruct $s$ since it is
smaller than the product of the smallest $t$ primes. As all the parameters
except the private shares $s_i$ are public, the secret reconstruction
is a straightforward application of CRT. It is
important to notice that the Mignotte $(t,n)$-threshold secret-sharing
scheme is not perfect in the sense that a set of less than $t$ shares
reveals some information about the secret.

\subsection{Asmuth-Bloom's secret sharing}
Let $p_0$ be a prime which defines the secret space and $s \in {\mathbb Z}_{p_0}$ be the secret. Let $M$ be $\prod_{i=1}^t
p_i$, and $p_0 < p_1 < p_2 < \ldots < p_n$ be an increasing sequence of primes
such that 
\begin{equation}
p_0 \prod_{i=1}^{t-1} p_{n-i+1} < M.
\label{eq:ab}
\end{equation}
\noindent To share the
secret, the dealer first chooses a random positive integer $\alpha$
such that $0 \le y = s + \alpha p_0 < M$. The share of the participant
${u_i}$ is equal to $s_i = y \,\, \mod \, p_i$. Let $A \in \mathcal{A}$ be a coalition
of $t$ participants and let 
$M_A=\prod_{i\in A} p_i$. Then the shared integer $y$ 
can be uniquely reconstructed in $\mathbb{Z}_{M_A}$ since $y <
M \leq M_A$. Hence, the secret $s$ can later be obtained by
computing $y \mod p_0$.

Asmuth Bloom's SSS has better security properties when compared to
Mignotte's. When a coalition $A'$ with $t-1$ shares tries to
reconstruct the secret, due to~\eqref{eq:ab}, there will be at least
$\frac{M}{M_{A'}} > p_0$ candidates for $y$. Furthermore,
since $p_0$ is relatively prime with $M_{A'}$, there will be
at least one $y$ candidate valid for each possible secret candidate in
${\mathbb Z}_{p_0}$. Thus, $t-1$ or fewer participants cannot narrow
down the secret space. However, since the number of $y$ candidates for
two secret candidates may differ~(by one), the secret candidates are
not equally probable, resulting in an imperfect
distribution~\cite{Kaya07}. To solve this problem,
Kaya~et~al. proposed to use the equation
\begin{equation}
{p_0}^2 \prod_{i=1}^{t-1} p_{n-i+1} < M
\label{eq:ab2}
\end{equation} 
instead of~\eqref{eq:ab}, which forms a \textit{statistical scheme} 
with respect to the definition given in \cite{Beimel11}. 
We will follow the same idea in this work.
\begin{table}[ht!]
\begin{center}
\begin{tabular}{ll}
\hline\noalign{\smallskip}
Notation&Explanation\\
\noalign{\smallskip}\hline\noalign{\smallskip}
$\mathcal{U}$& The set of participants.\\
$\mathcal{A}$& The collection of authorized subsets of $\mathcal{U}$, the access structure.\\
$n$&The number of total participants.\\
$m$&The number of levels$\backslash$compartments.\\
$u_k$&The k$^{th}$ participant.\\
$L_i$&The i$^{th}$ level$\backslash$compartment.\\
$n_i$&The number of participants in $L_i$.\\
$t_i$&The threshold, the minimum number of users\\ 
& required to construct the secret for level $L_i$.\\
$U_i$&$\sum_{k=1}^i L_k$.\\
$s$&The secret to be shared.\\
$s^j_k$&$y_j \bmod p_k$, the share of user $u_k\in L_j$.\\
$\Delta s^i_k$&$y_i- h_k(s^j_k,i) \bmod p_k$, the public information of user $u_k$ for $L_i$.\\
$M_{i}$&The modulus of smallest $t_i$ ones, $\prod_{j=1}^{t_i} {p_j}$.\\
$M_A$&The modulus of coalition $A$, $\prod_{u_i \in A}{p_i}$.\\
$p_0$&A prime; specifies the domain of $s \in {\mathbb Z}_{p_0}$.\\ 
$p_i$&The prime modulus for user $i$.\\ 
$y_i$& $s_i+\alpha_i\cdot p_0$, where $\alpha_i$ is the blinding factor.\\
\hline\noalign{\smallskip}
\end{tabular}
\caption{Notation}
\label{tab:not}
\end{center}	
\end{table}
For the rest of the paper, we will use the notation given in Table \ref{tab:not}.


\subsection{Multilevel threshold secret sharing}
We employ Simmons' multilevel threshold secret sharing~(MTSS)
definition, which assumes a multipartite access structure and a
hierarchy on it such that the members of the superior
compartments~(higher-level members) can replace the ones from inferior
compartments~(lower-level members). Throughout the paper, the terms
\emph{level} and \emph{compartment} are used interchangeably for our
context.

Let $\mathcal{U}$ be a set of all participants composed of disjoint
subsets called \textit{levels}, i.e, $\mathcal{U}= \bigcup _{i=1}^m L_i$
where $L_i \cap L_j = \emptyset$ for all $1 \le i,j \le m$. Here $L_1$
is the highest level and $L_m$ is the lowest one. Thus, a participant in
$L_1$ can take place of any other participant, and a participant in $L_m$
can only take place of the participants in $L_m$. Let the
integers $0
< t_1 < t_2 < \ldots < t_m$ be a sequence of threshold values such that $t_j \le
|L_1|+|L_2|+\ldots+|L_j|$ for all $ 1 \leq j \leq m$. When considered
in the disjunctive setting, the access structure is
defined by using the disjunction of the $m$ conditions on $m$
compartments as described below. 

\begin{defn} A $(t, n)$ disjunctive multilevel threshold secret sharing
scheme assigns each participant $u \in \mathcal{U}$ a secret share such
that the access structure is defined as $\mathcal{A} = \{A \subset \mathcal{U}: \exists i\in\{1,2,\ldots,m\} \mbox{ s.t. }|A \cap (\bigcup _{j=1}^i L_j )| \ge
t_i\}$. \end{defn}

\noindent On the other hand, under the conjunctive setting, all the
threshold conditions of the compartments need to be satisfied. We use
the same access structure definition as of \cite{Tassa04}.

\begin{defn} A $(t, n)$ conjunctive multilevel threshold secret sharing
scheme assigns each participant $u \in \mathcal{U}$ a secret share such
that the access structure is defined as $\mathcal{A} = \{A \subset \mathcal{U}:
\forall i\in\{1,2,\ldots,m\} \mbox{ s.t. } |A \cap (\bigcup _{j=1}^i L_j )| \ge
t_i\}$. \end{defn}

\subsection{The Harn-Fuyou MTSS scheme}

Assume that the participants are partitioned into $m$ levels $L_i$,
$i=1,2,\ldots,m$. Let $|L_i| = n_i$ be the number of participants in
$L_i$ and let $t_i < n_i$ define a threshold on it.  
The threshold of a higher-level is always smaller than the
threshold of a lower-level (i.e., $t_j < t_i$ for $j < i$) consistent
with the above MTSS definition. The disjunctive MTSS of Harn and Fuyou has two
phases:
\begin{itemize}

\interfootnotelinepenalty=10000
\item{\textit{{Share generation:}}
The dealer first selects a prime $p_0$, defining the secret space as 
$s \in \mathbb{Z}_{p_0}$. For each subset $L_i$ having $n_i$
participants, she selects a sequence of pairwise coprime positive
integers~(or primes), $p^i_1<p^i_2<\ldots<p^i_{n_i}$, such that $$p_0
p^i_{n_i-t_i+2} p^i_{n_i-t_i+3} \ldots p^i_{n_i} < p^i_1 p^i_2 \ldots
p^i_{t_i},$$ and $gcd(p_0,p^i_k)=1,k=1,2,\ldots,n_i$, where $p^i_k$ is
the public information associated with participant $u^i_k$, the
$k^{th}$ member of the subset $L_i$. For each such sequence, the
dealer selects an integer $\alpha_i$ such that the value
$s+\alpha_{i}p_0$ is in the $t_i-$\textit{threshold}
range~\cite{Harn14}. That is, $\alpha_i$ is chosen such that
$$p^i_{n_i-t_i+2} p^i_{n_i-t_i+3} \ldots p^i_{n_i} < s+\alpha_{i}p_0 <
p^i_1 p^i_2 \ldots p^i_{t_i}$$ supposedly in order to prevent the
recovery of the value $s+\alpha_{i}p_0$ with fewer than $t_i$
shares.\footnote{In the Harn-Fuyou scheme, the lower-bound on $y_i = s+\alpha_{i}p_0$
constitutes an extra restriction on the original Asmuth-Bloom scheme
and this range is called \textbf{t-threshold range} therein. That is,
while the upper bound $M_i = \prod_{j=1}^{t_i} p^i_j$ remains the same,
the lower bound that $y_i = s+\alpha_{i}p_0$ can attain is restricted
to be greater than $p^i_{n-t+2}\ldots p^i_{n_i}$ rather than $0$. Thus,
Harn-Fuyou employs a slightly different version of the Asmuth-Bloom
scheme. In our scheme, we will follow the original bounds.}

For each participant $u^i_k$, the private share $s^i_k$ that can directly be
used for level $L_i$ is generated as $s^i_k = s+\alpha_{i}p_0 \,\,
\mod \, p^i_k$. In order to enable the use of $s^i_k$ in a compartment
$L_j$~($j > i$), the dealer first selects a prime $p^i_{k,j}$ such that
$p^j_{t_j}<p^i_{k,j}<p^j_{n_j-t_j+2}$. She then computes $$\Delta
s^i_{k,j} = (s+\alpha_{j}p_0 - s^i_k) \, \mod \, p^i_{k,j}$$ and
broadcasts it with $p^i_{k,j}$ as a public information. 

All selected $p^i_{k,j}$s during this phase must be relatively coprime
to all other moduli. At the end of the phase, each participant
$u^i_k \in L_i$ keeps a single private share
$s^i_k \in \mathbb{Z}_{p^i_k}$ accompanied with the public information
$(\Delta s^i_{k,j} , p^i_{k,j})$ for $j \in \{i+1,i+2,\dots,m\}$.}

\item{\textit{{Secret reconstruction:}} 
The secret can be recovered by a coalition of participants if there
are at least $t_j$ participants in the coalition from levels $L_i$
where $1 \leq i \leq j.$ By using the corresponding shares, a system
of equations regarding CRT can be established on the joined shares; if
the participant $u^i_k$ belongs to $L_j$, i.e., $i = j$, she can use
her share $s^i_k$ and the modulus $p^i_k$ directly. Otherwise, i.e.,
if $i < j$, her share needs to be modified as $s^i_k+\Delta s^i_{k,j}$ to
be used in the lower level $L_j$ and the operations for this modified
share need to be performed in modulo $p^i_{k,j}$ while constructing the
system of CRT equations. Using all these shares and a standard CRT
construction, a unique solution $y = s + \alpha_j p_0$ can be
obtained. Then the secret can be reconstructed by computing $s = y \,
\mod \, p_0$.}
\end{itemize}

\section{The Fallacies of Harn-Fuyou MTSS Scheme}\label{sec:HY} Although
the Harn-Fuyou scheme employs interesting and useful mini-mechanisms
resulting in the first MTSS scheme employing CRT, there are some unresolved
issues as will be discussed here. A minor problem is that their
MTSS is based on the original Asmuth-Bloom scheme which is not
perfect~(i.e., the secret candidates are not statistically equally
likely to be the secret for an invalid coalition with $t-1$ shares). Although, this can be neglected if the secret
is shared only once, sharing the same secret multiple times with a
non-perfect scheme in practice may cause significant probabilistic differences in
the secret space. For that reason, we believe that instead of the
original scheme, the modified version proposed in~\cite{Kaya07} is more
appropriate for a MTSS scheme.

The proposed scheme is not generic since there are practical cases for
which it cannot be employed; as mentioned above, in the {\it share
generation} phase, there are additional $p^i_{k,j}$ values associated
with each participant $u^i_k$ for each level $L_j$ lower than hers. These
numbers need to fulfill the condition $p^j_{t_j} < p^i_{k,j} <
p^j_{n_j-t_j+2}$ and hence, the scheme implicitly compels the dealer to
initially select the primes $p^j_1 < p^j_2 < \ldots < p^j_{n_j}$ with a
gap allowing sufficient number of primes in between $p^j_{t_j}$ and
$p^j_{n_j-t_j+2}$ so that $p^i_{k,j}$s can fill in. In addition to the
gap, $p^j_{t_j} < p^i_{k,j} < p^j_{n_j-t_j+2}$ explicitly states that
$t_j < n_j-t_j+2$. Therefore, the Harn-Fuyou scheme is not suitable for
the cases where the compartment threshold composes at least one more than
the majority of the participants as the following simple setting shows.

\noindent \textbf{Example 1:}
Let there be two levels $L_1$ and $L_2$ involving
$n_1=|L_1|=2$ and $n_2=|L_2|=3$ participants and let the thresholds
be $t_1=2$ and $t_2=3$. The dealer selects the primes 
$p_0 < p^1_1 < p^1_2$ and $p_0 < p^2_1 < p^2_2 < p^2_3$ which need to satisfy
$$p_0p^1_2 < p^1_1p^1_2$$ $$p_0p^2_2p^2_3<p^2_1p^2_2p^2_3$$ to be secure. Recall that $p^i_{k,j}$ is the
prime distributed to $k^{th}$ user in $i^{th}$ level to be used for
participation in a lower compartment $j$. Since $p^i_{k,j}$ must be
chosen such that $p^j_{t_j}<p^i_{k,j}<p^j_{n_j-t_j+2}$,  we have
$p^2_{3} < p^1_{1,2} < p^2_{2}$ and $p^2_{3} < p^2_{2}$
contradicts with the initial choice of primes $p^2_{2} <
p^2_{3}$.

Hence, placing the primes $p^i_{k,j}$ between $p^j_{t_j}$ and
$p^j_{n_j-t_j+2}$ requires a condition which is not guaranteed to hold
in a generic setting; it simply may be the case that $p^j_{t_j} >
p^j_{n_j-t_j+2}$, i.e., $t_j > \lceil\frac{n_j}{2}\rceil+1$. That is,
the existence of some interval in between the primes is not ensured
since there is no order whatsoever among the primes of different
compartments.

The most important problem of the Harn-Fuyou scheme is in fact its mismatch 
with the multilevel access structure of Simmons. In
general, the range of the threshold values $t_i$ are given such as $1
\le t_i \le \sum_{j=1}^i |L_j|$ for $i=1,2,\ldots,m$. Hence,
$t_i$ can be greater than $n_i=|L_i|$ as $\sum_{j=1}^i
|L_j|>|L_i|$. Nonetheless, in the Harn-Fuyou scheme, the specified
primes $p^i_1< p^i_2< \ldots, <p^i_{n_i}$ cease at the index $n_i$,
resulting in the condition $p_0 p^i_{n_i-t_i+2} p^i_{n_i-t_i+3}\ldots p^i_{n_i} <
p^i_1 p^i_2 \ldots p^i_{t_i}$ being unclear for large enough $t_i$ that
exceeds $n_i$. 
For example, the scheme is not well-defined for a setting
with two compartments $L_1$ and $L_2$, where $n_1=3$, $n_2=3$, $t_1=2$ and $t_2=4$ since there are only $3$ users
in the second compartment. The 
threshold is $4$ and a $(t,n)$-Asmuth-Bloom sequence with $n=3$ and $t=4$ does not exist.

\subsection{A straightforward (yet insecure) modification of the Harn-Fuyou MTSS}

One can make the Harn-Fuyou MTSS scheme suitable for any number of
participants and threshold values by removing the necessity of the additional primes: In the share generation
phase, instead of using a sequence with $n_i$ primes $p^i_1 < p^i_2 <
\ldots <p^i_{n_i}$ for compartment $L_i$, the dealer can use a sequence
with ${U_i}$ primes $p^i_1 < p^i_2 <
\ldots <p^i_{U_i}$ where $U_i=\sum_{j=1}^i n_j$. For security, the
condition to be satisfied for this prime set is
$$p_0 p^i_{U_i-t_i+2} p^i_{U_i-t_i+3} \ldots p^i_{U_i} < p^i_1 p^i_2 \ldots p^i_{t_i}$$
%
that is well defined for any valid value of $t_i$. 
Here, the first $n_i$ primes can be used for the participants in
$L_i$ and the extra primes $p^i_{\ell}$ for $\ell > n_i$ can be used
for $p^i_{k,j}$s for the participants in higher compartments. The random
integers $\alpha_i, 1\le i\le m$ are chosen such that $0 \le s+\alpha_i
p_0 < p^i_1p^i_2 \ldots p^i_{t_i}$. The share $s^i_k$ for participant
$u^i_k$ is generated as $s^i_k = s+\alpha_{i}p_0 \, \mod \, p^i_{k,j}$
as before.

This approach indeed eliminates the need for $p^i_{k,j}$ to fill in to a
possibly non-existing gap in between  $p^j_{t_j} < p^i_{k,j} <
p^j_{n_j-t_j+2}$. As this is the only distinction we describe herein,
the rest of the share generation phase and the secret reconstruction
phase remains essentially intact, and can be performed  in a similar
fashion as described before. 

Although we established a well-defined scheme for all possible threshold settings, this approach
unfortunately does not provide security as the following example
illustrates. The example below is given for the modified/fixed version
without the {\it gap existence} problem. However, the weakness also
exists in the original MTSS scheme of Harn-Fuyou since the public 
information with different prime modulos for a certain participant
reveals extra information as we will show below. 

\noindent \textbf{Example 2:} Consider the following setting emerging
from the scheme with the basic fix above. Let $p_0=5$ and $s = 1 \in
\mathbb{Z}_{5}$. Suppose that we have two compartments $L_1$ and $L_2$
with $n_1=4$, $n_2=2$, $t_1=2$ and $t_2=3$. 
Let $$p^1_1 < p^1_2 < p^1_3 <
p^1_4 \rightarrow 11 < 13 < 17 < 23$$ 
$$p^2_1 < p^2_2 < p^2_3 < p^2_4 < p^2_5 < p^2_6 \rightarrow 
29 < 31 < 37 < 61 < 67 < 71$$
be the primes. The Asmuth-Bloom condition $p_0 p^i_{U_i-t_i+2} p^i_{U_i-t_i+3} \ldots
p^i_{U_i}  < p^i_1 p^i_2 \ldots p^i_{t_i}$ is satisfied for both levels since 
$$5\times23 = 115 < 143 = 11\times13,$$ $$5\times67\times71 = 23785 < 33263 = 29\times31\times37.$$ 

\noindent Let $\alpha_1=5$ and $\alpha_2=952$. Hence, $$y_1 = s+\alpha_1p_0 =
1+5\times5=26,$$  $$y_2 = s+\alpha_2p_0 =
1+952\times5=4761,$$ and these values are chosen from the t-threshold range  since
$$23 < 26 < 143 = 11\times13,$$  $$67 \times 71   = 4757 < 4761 <
33263 = 29\times31\times37.$$

\noindent Similarly, let  
$p^1_{1,2}=p^2_6=71,\ \ 
p^1_{2,2}=p^2_5=67,\ \ 
p^1_{3,2}=p^2_4=61,\ \ 
p^1_{4,2}=p^2_3=37$
be the additional primes that will be used
to enable the share of the participants in $L_1$ for $L_2$. With these parameters, the shares are    
$$s^1_1 = 4, s^1_2 = 0, s^1_3=9, \mbox{ and } s^1_4= 3,$$ $$s^2_1=5 \mbox{\  and\  } 
s^2_2=18$$ $$s^1_{1,2}=25, s^1_{2,2}=3, s^1_{3,2}=4, \mbox{\  and\  }  s^1_{4,2}=4,$$ and the public information
is computed as $$\Delta s^1_{1,2}=0, \Delta
s^1_{2,1}=4, \Delta s^1_{3,1}=55, \Delta s^1_{4,1}=22.$$ To exemplify, the
first participant in $L_1$ is associated with the prime $p^1_1=11$ as
well as the prime $ p^1_{1,2}=71$ and the integer $\Delta
s^1_{1,2}=(s+\alpha_2p_0)-s^1_1 = 0$ (in $\bmod\
 p^1_{1,2}$).

Suppose that the adversary corrupted $u^2_1$ and $u^2_2$ hence obtained
their shares. She knows that $y_2$ is  bounded by $4757 < y_2 < 33263$ and she
also can compute $y_2 \mod p^2_1p^2_2 = y_2 \mod 899 = 266$ by using
these shares. There are $\lceil(33263 - 4757) / 899\rceil = 32$ candidates for $y_2$ all in
form $266 + 899\times K$ where $5 \leq K \leq 36$. Since, $899$ is
relatively prime with $5$, each secret candidate in $\mathbb{Z}_{p_0}$
must be valid for around  $7$ of these values, i.e., for $266 + 899\times 6$ the valid secret candidate is $0$. 
Hence, without the public information, thanks to the perfectness of Asmuth-Bloom SSS, the adversary cannot have an information on the secret. 
Unfortunately, with public information that reveal extra information for the
uncorrupted users, the adversary can gain some information on the secret and sometimes she even can 
find the secret as shown in the example below.

The participant $u^1_1$ has a public
information pair $(\Delta s^1_{1,2}, p^1_{1,2}) = (0, 71)$ and her
prime is $p^1_1 = 11$. Hence, the adversary knows that the value
$s^1_{1,2}$ is bounded by $s^1_{1,2} = s^1_1 + \Delta s^1_{1,2} \in
[0, 10]$ since $s^1_1 \in \mathbb{Z}_{11}$. Similarly, for $u^1_2$,
$u^1_3$, and $u^1_4$, the adversary knows that $$s^1_{2,2} \in [4, 16],
s^1_{3,2} \in [55, 60] \cup [0,10], s^1_{2,2} \in [22,36] \cup [0,7].$$ As the
Table~\ref{tab:ex3} shows, there is only one $y_2$ candidate in the form
$51+899×K$, which yields $s^1_{\{1,2,3,4\},2}$ values within these ranges. 
Thus the adversary knows that $y_2 = 4761$ and the secret $s=1$ is recovered in an unauthorized manner by 
corrupting only two participants from $L_2$.

\renewcommand{\arraystretch}{0.6}
\begin{table*}
\center
\begin{tabular}{r|rrrr||r|rrrr}
candidate & $s^1_{1,2}$ & $s^1_{2,2}$ & $s^1_{3,2}$ & $s^1_{4,2}$ & candidate & $s^1_{1,2}$ & $s^1_{2,2}$ & $s^1_{3,2}$ & $s^1_{4,2}$\\\hline
4761 & \bf{4} & \bf{4} & \bf{3} & \bf{25} & 19145 & 46 & 50 & 52 & 16\\
5660 & 51 & 32 & 48 & \bf{36} & 20044 & 22 & \bf{11} & 36 & \bf{27}\\
6559 & 27 & 60 & 32 & 10 & 20943 & 69 & 39 & 20 & \bf{1}\\
7458 & \bf{3} & 21 & 16 & 21 & 21842 & 45 & 0 & \bf{4} & 12\\
8357 & 50 & 49 & \bf{0} & \bf{32} & 22741 & 21 & 28 & 49 & \bf{23}\\
9256 & 26 & \bf{10} & 45 & \bf{6} & 23640 & 68 & 56 & 33 & \bf{34}\\
10155 & \bf{2} & 38 & 29 & 17 & 24539 & 44 & 17 & 17 & 8\\
11054 & 49 & 66 & 13 & \bf{28} & 25438 & 20 & 45 & \bf{1} & 19\\
11953 & 25 & 27 & \bf{58} & \bf{2} & 26337 & 67 & \bf{6} & 46 & \bf{30}\\
12852 & \bf{1} & 55 & 42 & 13 & 27236 & 43 & 34 & 30 & \bf{4}\\
13751 & 48 & 16 & 26 & \bf{24} & 28135 & 19 & 62 & 14 & 15\\
14650 & 24 & 44 & \bf{10} & \bf{35} & 29034 & 66 & 23 & \bf{59} & \bf{26}\\
15549 & \bf{0} & \bf{5} & \bf{55} & 9 & 29933 & 42 & 51 & 43 & \bf{0}\\
16448 & 47 & 33 & 39 & 20 & 30832 & 18 & \bf{12} & 27 & 11\\
17347 & 23 & 61 & 23 & \bf{31} & 31731 & 65 & 40 & 11 & \bf{22}\\
18246 & 70 & 22 & \bf{7} & \bf{5} & 32630 & 41 & 1 & \bf{56} & \bf{33}\\
\end{tabular}
\caption{Secrets for each $y_2$ candidate from adversary's point of view for Example 2.\newline The values consistent with the ranges obtained by public information are shown in boldface.} 
\label{tab:ex3}
\end{table*}

\section{Proposed Multilevel Threshold Secret Sharing Schemes}\label{sec:MSSS}

As described before, we are given a secret $s \in {\mathbb Z}_{p_0}$ and
a set of primes such that
\begin{equation}
{p_0}^2 \prod_{i=1}^{t-1} p_{n-i+1} < \prod_{i=1}^t p_i,
\label{eq:AB}
\end{equation}
i.e., the \textit{Asmuth-Bloom condition} holds. We will refer to the prime
sequence $p_0 < p_1 < p_2 < \ldots < p_n$ satisfying the Asmuth-Bloom condition as a
\textit{$(t,n)$-Asmuth-Bloom sequence}. As the fallacies of the
Harn-Fuyou scheme show, having the Asmuth-Bloom condition for all the
compartments independently while keeping the level structure and being secure is not an
easy task. We solve this problem by using a single \textit{anchor
Asmuth-Bloom sequence} as defined below so that each participant of the
MTSS has only one prime modulus that can be used for all the levels she
can contribute to. 

\begin{defn} An anchor Asmuth-Bloom sequence is a sequence of primes
$p_0 < p_1 < p_2 < \ldots < p_n$ satisfying 
\begin{equation}
\displaystyle {p_0}^2
\prod_{i=1}^{\left \lfloor{n/2}\right \rfloor-1} p_{n-i+1} < 
\prod_{i=1}^{\left \lfloor{n/2}\right \rfloor} p_{i}.
\label{eq:anchor}
\end{equation}
\end{defn}

\noindent As one can notice, an anchor sequence is a valid $(\left
\lfloor{n/2}\right \rfloor, n)$-Asmuth-Bloom sequence. Here, we will show
that, an anchor sequence can be used not only for $t = \left
\lfloor{n/2}\right \rfloor$ but also for other $t$ values:

\begin{lem} An anchor Asmuth-Bloom sequence can be employed for any
CRT-based $(t, n)$ secret sharing scheme. That is an anchor prime sequence
satisfies the Asmuth-Bloom condition for any $1 \le t \le
n$.\end{lem}

\noindent \textbf{Proof:} We will investigate the sequence in two cases:
\begin{enumerate}

\item ($t<\left \lfloor{n/2}\right \rfloor$): To
have~\eqref{eq:AB} from~\eqref{eq:anchor} for a threshold value $t<\left
\lfloor{n/2}\right \rfloor$, one can remove $\left \lfloor{n/2}\right
\rfloor - t$ primes from each side of~\eqref{eq:anchor}. Note that for
each prime $p_i$ removed from the right side, one needs to remove
$p_{n-i+1}$ from the left. Since $i \leq t  < \left \lfloor{n/2}\right \rfloor$
for all the primes removed, $n-i+1 > i$ and $p_{n-i+1} > p_i$. Thus, given
the anchor inequality~\eqref{eq:anchor}, the Asmuth-Bloom condition~\eqref{eq:AB} is also satisfied for 
a threshold $t<\left\lfloor{n/2}\right \rfloor$ with the same set of primes.

\item ($t \ge \left \lfloor{n/2}\right \rfloor$): This case is similar to
the former case except that to have~\eqref{eq:AB}
from~\eqref{eq:anchor}, we need to add $t - \left \lfloor{n/2}\right
\rfloor$ primes to each side of~\eqref{eq:anchor}. For each prime pair
$(p_{n-i+1}, p_i)$ added to the left and right of the anchor inequality,
respectively, $p_{n-i+1} < p_i$ since $t  \leq i > \left \lfloor{n/2}\right
\rfloor$. Thus given~\eqref{eq:anchor},~\eqref{eq:AB} is also satisfied
for a threshold value $t \geq \left\lfloor{n/2}\right \rfloor$ with the same prime sequence. $\blacksquare$

\end{enumerate}
\subsection{A novel CRT-based multilevel threshold (disjunctive) SSS}

Let $n = \sum_{i = 1}^m n_i$ be the number of total participants. Let
$h_i: \mathbb{Z}_{p_i} \times \mathbb{Z}_{m} \rightarrow
\mathbb{Z}_{p_i}$ for $i \in \{1, \ldots, n\}$ be a family of
efficiently computable one-way hash functions. We employ an anchor sequence of $n$ primes as follows:

\begin{itemize} 

\item{\textit{Initialization:}} The dealer first generates an anchor
prime sequence $p_0 < p_1 < p_2 < \ldots < p_n$
satisfying~\eqref{eq:anchor} and assign each prime $p_i$ to a
participant $u_i$. Note that this will be the
only prime modulus that will be used for the participant\footnote{While describing the proposed schemes, we will denote the primes and participants with a
single subscript as opposed to the notation in Harn-Fuyou scheme. We believe this is more clear thanks to the compactness of the anchor sequence we employ.}.

\item{\textit{Share generation:}} Given a secret $s \in {\mathbb
Z}_{p_0}$, the dealer chooses $\alpha_i$'s for all $1 \le i \le m$ such
that $$0 \le y_i = s + \alpha_i p_0 < M_i = p_{1}p_{2} \ldots p_{t_i}.$$ 

For level $L_i$, the shares and the public information are generated as
follows: Let $u_k$ be a participant in $L_i$; the original share $s^i_k$
for $u_k$ is generated as $s^i_k = y_i \mod p_k$. 

If $u_k$ is a participant in a higher compartment $L_j$, i.e., $j < i$;
to enable the use of $s^j_k$ in $L_i$, the dealer computes $\Delta
s^i_{k} = (y_i - h_k(s^j_k, i)) \mod p_k$ and broadcasts it as the
public information. This information will be used if $u_k$ participates
in the secret reconstruction within $L_i$. 

\item{\textit{Secret reconstruction:}} Let $A$ be a coalition
gathered to reconstruct the secret. $A$ is an authorized
coalition if it has $t_i$ or more participants from $L_i$ or higher
compartments for $1 \le i \le m$. If the participant is from $L_i$, her
share $s^i_k$ can be used as is. Any other share $s_{k}^j$ of $u_k$ from
a higher level needs to be modified as $(s_{k}^j + \Delta s_{k}^i)$ and
is used with the modulus $p_k$ while constructing the system of congruences.
Using the standard CRT, a unique solution $y_i$ can be obtained. Then,
the secret $s$ is recovered by computing $s = y_i \mod p_0$.

\end{itemize}

An authorized coalition can obtain the secret since with the help of
public information, the coalition will have enough shares for a
compartment $L_i$. Thanks to CRT, the corresponding $y_i$ value and
hence $s = y_i \bmod p_0$ can be computed.

\subsubsection{Security analysis of the proposed MTSS}
The security of the proposed MTSS solely depends on the security of the
Asmuth-Bloom scheme. We will argue that unlike the Harn-Fuyou scheme,
the proposed MTSS scheme does not reveal any information on the secret with the
public information used. Then, we will prove that an adversarial coalition
cannot have any information on the secret. 

To generate the public information, the proposed MTSS scheme employs a hash
function for each participant. Let $u_k$ be a participant in $L_j$. If
the adversary corrupts $u_k$ she will have $s^j_k$ and she can compute
the shares for all levels $L_j$, $j \leq i \leq m$. If $u_k$ remains
uncorrupted, the adversary will only have the public information for
$u_k$. Let $L_i$ be a level lower than $j$;
the adversary will have 
\begin{align} 
\Delta s^i_{k} &= (s^i_{k} - h_k(s^j_k, i)) \mod p_k 
\end{align} 
Hence, assuming the hash function $h_k$ behaves like a random oracle,
$\Delta s^i_{k}$ will be random~(which can be randomly generated in a
zero-knowledge proof). Thus the adversary cannot learn anything on the
shares of $u_k$ for lower compartments. Furthermore, although the same
hash function $h_k$ is used to compute $\Delta s^i_{k}$ and $\Delta
s^{i'}_{k}$ for two lower levels $L_i$ and $L_{i'}$, $j \leq i, i' \leq
m$, these two values cannot be combined~(as they could be without the
hash function), since $h_k$ takes $i$ and $i'$, respectively, as an
input. 

\begin{thm}\label{thm:sec_MTSS}
Given that the hash functions used in the MTSS scheme behave like random oracles, 
an unauthorized coalition cannot obtain any information about the secret.  
\end{thm}

\noindent \textbf{Proof:} Let $A'$ be the adversarial
coalition having $t_i - 1$ participants from $L_i$ and higher
compartments. Let $M_{A'}$ be the product of the prime modulus
values assigned to these $t_i - 1$ participants and $y'_i = y_i \mod
M_{A'}$. Since ${p_0}^2\prod_{j=1}^{t_i-1} p_{n-j+1} <
\prod_{j=1}^{t_i} p_j < \prod_{j=1}^{t_i} {p_j}= M_i$, we have $M_i / M_{A'} > {p_0}^2$. Hence
$y'_i + \beta M_{A'}$ is a valid candidate for $y_i < M$ for
all $\beta < {p_0}^2$. Since $gcd(p_0,M_{A'}) = 1$, all $(y' +
\beta M_{A'}) \mod p_0$ are distinct for $\ell p_0 \le \beta <
(\ell+1)p_0$, for each $0 \leq \ell < p_0$. Thus $s$ can be any integer from
$\mathbb{Z}_{p_0}$ and the secret space is not restricted from 
the adversary's point of view.

For each value $s'$ in the secret space, from the adversary's point of
view, there are either ${\left \lfloor{M_i/(M_{A'}p_0)}\right
\rfloor}$  or ${\left \lfloor{M_i/(M_{A'}p_0)}\right \rfloor}
+ 1 $ possible consistent $y_i$ candidates consistent with $s'$.
Considering $M_i/M_{A'} > {p_0}^2$, for two different
integers $s'$ and $s''$ in ${\mathbb{Z}_{p_0}}$, the probabilities of $s
= s'$ or $s = s''$ are almost equal and the difference between these
two values reduces when $p_0$ increases. More formally, thanks to the modified
Asmuth-Bloom SSS we employed \cite{Kaya07}, the proposed MTSS scheme is {\it
statistical}, i.e., the statistical distance between the probability
distribution of the secret candidates being a secret and an uniform
distribution is smaller than a given $\epsilon$ with a carefully chosen
$p_0$. $\blacksquare$

\subsection{A CRT-based multilevel threshold (conjunctive) SSS}

The ideas presented above for the disjunctive scheme can also be
employed to have a conjuctive SSS. Here, we present the first CRT-based
conjunctive MTSS scheme which adopts Iftene's CRT-based compartmented SSS
~\cite{Iftene05}. 

The setting is the same as that of the disjunctive MTSS scheme; compartment
$L_i$ with threshold $t_i$ has $n_i$ participants for $1 \leq i \leq m$.
Hence, the total number of participants is $n = \sum_{i=1}^{m} n_i$.
There is a hierarchy between the compartments; a
member of $L_j$ can act as a member of a lower compartment $L_i$ if $i >
j$. The proposed conjunctive scheme shares a given secret $s \in
\mathbb{Z}_{p_0}$ as follows:

\begin{itemize}

\item{\textit{Initialization:}} The anchor prime sequence generation is the
same. Let $\sigma_1, \sigma_2, \ldots, \sigma_{m-1}$ be random integers from 
$\mathbb{Z}_{p_0}$ and $\sigma_m \in \mathbb{Z}_{p_0}$ is chosen such that
$$s = (\sigma_1 + \sigma_2 + \cdots + \sigma_m) \bmod p_0 .$$

\item{\textit{Share generation:}}  For all $1 \leq i \leq m$, a random
$\alpha_i$ is chosen such that $0 \le y_i = \sigma_i + \alpha_i p_0 < M_i =
 p_{1}p_{2} \ldots p_{t_i}$. The shares and public information are generated
similar to the disjunctive case. Let $u_k$ be a participant in $L_i$;
the original share $s^i_k$ for $u_k$ is generated as $s^i_k = y_i \mod
p_k$. For all $u_k$ who is from a higher level $L_j$ to enable the use of $s^j_k$
in $L_i$, $\Delta s^i_{k} = (y_i - h_k(s^j_k, i)) \mod p_k$ is computed
and broadcasted.

\item{\textit{Secret reconstruction:}} The secret $s$ can be recovered
if and only if all of the $\sigma_i$ values for $1 \leq i \leq m$ are
recovered. A partial secret $\sigma_i$ can be recovered if the number of
shares from level $L_i$ or from higher levels is greater than or equal
to $t_i$. Let $u_k$ be a coalition member participating in this task; if
$u_k \in L_i$, her original share $s_{k}^i$ can be used. Otherwise, if
$u_k \in L_j$ for $j < i$, $s_{k}^j + \Delta s_{k}^i$ is computed and
used as $s_{k}^i$. After computing all $\sigma_i$ values for $1 \leq i \leq m$, the
secret $s$ is constructed by $s = (\sigma_1 + \sigma_2 + \cdots + \sigma_m) \bmod p_0$.
\end{itemize}

Since the scheme uses exactly the same set of public information and the
underlying statistical SSS is the same, the security analysis for the
disjunctive case can also be applied for the conjunctive MTSS scheme with minor
modifications and is omitted here.

\section{Multilevel Threshold Function Sharing}\label{sec:MFSS}

In this section, we adapt our MTSS scheme to have a CRT-based multilevel
function sharing scheme~(FSS) which can be used for decrypting a
ciphertext or signing a message in a way that no unqualified coalition
of participants can perform this operation. Another important property
of a FSS is that it does not disclose the secret and the shares; thus,
it can be used several times without any rearrangement. Several
protocols for function sharing~\cite{Desmedt90,DeSantis94,Gennaro01}
have been proposed in the literature where most of them are based on the
Shamir SSS. Kaya and Sel\c{c}uk~\cite{Kaya07} proposed the first
CRT-based FSS for RSA signature~\cite{RSA} and the
ElGamal~\cite{ElGamal85} decryption functions. Here, we propose a secure
CRT-based multilevel function sharing scheme~(MFSS) for RSA signatures
to demonstrate that the proposed MTSS scheme is applicable for function sharing.


\paragraph*{A Threshold (disjunctive) RSA Signature Scheme}

Let $N=pq$ be product of two large primes. Choose public key $e$ and
private key $d$ such that $ed\equiv 1 \pmod{\phi(N)}$. The signature of
a message $\mathbf{msg}$ is $\mathbf{sgn} = \mathbf{msg}^d\pmod N$ and the verification is done by
checking $\mathbf{msg}\stackrel{?}{=}\mathbf{sgn}^e\pmod N$. The setup phase for the
threshold multilevel RSA scheme is given in Figure~\ref{fig:MTFS1} and
the signature and verification steps can be found in
Figure~\ref{fig:MTFS2}. Here we describe a disjunctive scheme but the
scheme can be converted to the conjunctive case with minor
modifications. 

\begin{figure*}[ht!]
\begin{center}
\small
\framebox[5.5in]{\begin{minipage}[t]{5.4in}
\begin{algorithmic}
\STATE - Let $N=pq$ be product of two strong primes, i.e. $p=2p'+1$ and $q=2q'+1$ where $p'$ and $q'$ are large primes.
\STATE - Choose $e$ and $d$ such that $ed\equiv 1 \pmod{\phi(N)}$, ($\phi(N)=4p'q'$).
\STATE - Use the proposed MTSS scheme in order to share the secret $s = d$ with $p_0=4p'q'$ to $m$ levels where the $i^{th}$ level $L_i$ has
$n_i$ users with a threshold $t_i$.
\end{algorithmic}
\end{minipage}}
\end{center}
\caption{Setup of the proposed multilevel (disjunctive) threshold RSA signature scheme.}\label{fig:MTFS1}
\end{figure*}

\renewcommand{\baselinestretch}{0.9}
\begin{figure*}[ht!]
\begin{center}
\small
\framebox[5.5in]{\begin{minipage}[t]{5.3in}
\begin{algorithmic}
\STATE
\STATE \underline{Signing}
\STATE - Let $\mathbf{msg}\in \mathbb{Z^*_N}$ be the message to be signed.
\STATE - Let $A\in \mathcal{A}$ be a coalition in the access structure wants to sign $\mathbf{msg}$.
\STATE - Let $i$ be an integer s.t. $A_i = A \cap (\bigcup _{j=1}^i L_j)$ and $|A_i| \ge
t_i$.
\STATE - Each user $u_k \in A_i$ computes $$M_{A_i} = \prod_{u_{k'} \in A_i}p_{k'} \mbox{ and } P_k = \frac{M_{A_i}}{p_k} \bmod p_k$$
and $I_k$, the inverse of $P_k$ s.t. $I_kP_k \equiv 1 \bmod p_k$. She then computes the partial signature $\mathbf{sgn}_k$ as
$$\overline{s^i_k} =
\left\{
\begin{array}{ll}
s^i_k & \mbox{if } u_k \in L_i \\
s_{k}^j& \mbox{if } u_k \in L_j \mbox{ and } j < i
\end{array}
\right.$$
\vspace*{-3ex}
\begin{eqnarray*}
\nu_k&=& \overline{s_k^i} P_k I_k \bmod M_{A_i},\\
\mathbf{sgn}_k &=& \mathbf{msg}^{\nu_k} \bmod N
\end{eqnarray*}
and sends $\mathbf{sgn}_k$ to the server. 
\STATE - For each user  $u_k$, Server computes public parts of the signature as 
$$\overline{\Delta s^i_k} =
\left\{
\begin{array}{ll}
0 & \mbox{if } u_k \in L_i \\
\Delta s_{k}^i& \mbox{if } u_k \in L_j \mbox{ and } j < i
\end{array}
\right.$$
\vspace*{-3ex}
\begin{eqnarray*}
\Delta\nu_k&=& \Delta\overline{s_k^i} P_k I_k \bmod M_{A_i},\\	
\Delta\mathbf{sgn}_k &=& \mathbf{msg}^{\Delta\nu_k} \bmod N	
\end{eqnarray*}
\STATE Server combines all parts and computes incomplete signature $\overline{\mathbf{sgn}}$
\begin{eqnarray*}
\overline{\mathbf{sgn}}=\prod_{u_{k'} \in A_i} \left(\mathbf{sgn}_k\times\Delta\mathbf{sgn}_k\right)
\end{eqnarray*}
\STATE Server converts incomplete signature $\overline{\mathbf{sgn}}$ to the signature $\mathbf{sgn}$ by trying $x$
\begin{eqnarray}\label{eqn:MTSS}
(\overline{\mathbf{sgn}}\times \kappa^{x})^e \stackrel{?}{\equiv} \mathbf{msg} \bmod N
\end{eqnarray}
for $0\leq x<2|G|$ and let $\delta$ denotes the value of $x$ satisfying (\ref{eqn:MTSS}).
\STATE Then the signature is computed as $\mathbf{sgn}=\overline{\mathbf{sgn}}\times\kappa^{\delta}$
\end{algorithmic}
\end{minipage}}
\end{center}
\caption{Proposed multilevel (disjunctive) threshold RSA signature scheme: verification phase is not given since it is the same as the RSA verification}\label{fig:MTFS2}
\end{figure*}
\renewcommand{\baselinestretch}{1}

\paragraph*{Security Analysis} Since the proposed MTSS scheme is as secure as the
original Asmuth Bloom SSS by Theorem~\ref{thm:sec_MTSS} and the adapted
threshold signature scheme is proven to be secure with the Asmuth-Bloom
structure~\cite{Kaya07}, the proposed MFSS is also secure under the
assumption of intractability of the RSA problem. Detailed explanations
about random oracle proofs for the CRT-based threshold RSA can be found
in~\cite{Kaya07}.

\section{Conclusion}\label{sec:conc}
The CRT-based multilevel threshold SSS of Harn-Fuyou in the literature cannot be used for 
all threshold settings. Furthermore, the scheme is not secure and an adversary can extract the secret by using the private shares of the participants she corrupted and information revealed to the public during the secret sharing phase. We proposed novel, compact, and elegant disjunctive and conjunctive multilevel SSSs based on a special prime sequence called anchor sequence. We showed that the proposed schemes can easily be adopted for function sharing schemes which have numerous applications in applied cryptography.

\bibliographystyle{ieeetr}
\bibliography{references}
\newpage
\begin{IEEEbiographynophoto}{O\u{g}uzhan Ersoy}
received his B.S. degrees in Electrical \& Electronics Engineering and Mathematics from  Bo\u{g}azi\c{c}i University, \.{I}stanbul, Turkey, in 2012. He received his M.S. degree in 2015 in the Department of Electrical \& Electronics Engineering at Bo\u{g}azi\c{c}i University. He is working at T\"{U}B\.{I}TAK B\.{I}LGEM, Kocaeli, Turkey since 2012. 
\end{IEEEbiographynophoto}

\begin{IEEEbiographynophoto}{Kamer Kaya}
is working as an Assistant Professor at the Faculty of Engineering and Natural Sciences at Sabancı University, Turkey. He got his PhD from Dept. of Computer Engineering at Bilkent University, Turkey. His current research interests are Parallel Programming, Cryptography, and High Performance Computing.
\end{IEEEbiographynophoto}

\begin{IEEEbiographynophoto}{Kerem Ka{\c{s}}kalo{\u{g}}lu}
is working as an Assistant Professor at the department of Mathematics and Statistics at the American University of the Middle East, Kuwait. He received his PhD from the Middle East Technical University, Turkey. His current research interests are combinatorics and cryptography.
\end{IEEEbiographynophoto}

\end{document}